\documentclass[12pt,fleqn]{article}
\input{epsf}
\usepackage{amsmath}
\usepackage{latexsym}
\newlength{\dinwidth}
\newlength{\dinmargin}
\renewcommand{\thefootnote}{\dag}
\newcommand{\resection}[1]{\setcounter{equation}{0}\section{#1}}

\newcommand{\f}[2]{\frac{#1}{#2}}

\newcommand{\nn}{\nonumber}

\def\ap{\alpha^\prime}
\def\kkk{{\bf k}_\perp}
\def\ppp{{\bf p}^\prime_\perp}
\def\d{\partial}
\def\eps{\epsilon}
\def\pts{(\phi^3)_6}
\def\xb{{\bar x}}
\def\p{\phi}
\def\l{\f{\lambda^2}{(4\pi)^3}}
\def\mq{\left(\f{4\pi\mu^2}{Q^2}\right)^\epsilon}
\def\mt{\left(\f{4\pi\mu^2}{t}\right)^\epsilon}
\def\cc{\f{\lambda^2}{t^2}x} 
\begin{document}
 \thispagestyle{empty}

\begin{flushright}
 UPRF-97-09\\
\end{flushright}
\vspace*{1 cm}
 \begin{center}
 {\Large \bf One particle cross section in the target fragmentation region: an explicit calculation in $\pts$}\\
 \vspace*{1.5 cm}
 {\bf M. Grazzini}\footnote{Email: grazzini@parma.infn.it} \\
 {\em Dipartimento di Fisica, Universit\`a di Parma}\\
 {\em and INFN, Gruppo Collegato di Parma}
 \end{center}

\renewcommand{\thefootnote}{\arabic{footnote}}
\setcounter{footnote}{0}

 \begin{abstract}
 One particle inclusive cross sections in the target fragmentation region
 are considered and an explicit calculation is performed in $\pts$
 model field theory. The collinear divergences can be correctly absorbed
 into a parton density and a fragmentation function but the renormalized cross
 section gets a large logarithmic correction as expected in a two scale
 regime. We find that the coefficient of such a correction is precisely
 the scalar DGLAP kernel. Furthermore the consistency of this result
 with an extended factorization hypothesis is investigated.
 \end{abstract}
\vspace{.5cm}
\centerline{PACS 13.85.Ni}
\vspace*{2cm}

\newpage
\setcounter{page}{1}

\resection{Introduction}

Asymptotic freedom has
led to
important predictions for hard
inclusive hadronic processes in perturbative QCD.
The basic tools are factorization theorems \cite{fattorizzazione}
together with perturbative
evaluation of scaling properties.
Nevertheless perturbative QCD has been successfully
applied also to various semi-inclusive processes, for which a
complete factorization proof is still lacking.

In a deep inelastic semi-inclusive reaction a hadron with momentum $p$
is scattered by a far off shell photon with momentum $q$ ($Q^2=-q^2\gg
\Lambda_{QCD}^2$)
and from the inclusive final state, a hadron with momentum $p^\prime$  
is detected.
Until
some years ago
the main interest has been devoted
in hadron
production in the {\em current fragmentation region}, that is in the region
in which
the momentum transfer
$t=-(p-p^\prime)^2$
is of order $Q^2$ \cite{aemp}.
In the last few years,
after the introduction of {\em fracture functions} \cite{tv},
a novel attention has
been payed to the
{\em target fragmentation region} ($t\ll Q^2$) \cite{tv,grau,florian}.
With the observation of diffractive deep inelastic events at HERA
part of this attention has been focused on the limit
in which the longitudinal momentum fraction of the observed hadron
is close to unity and
$t$ is of order $1~GeV^2$ or less.

In this paper we want to deal with semi-inclusive deep inelastic scattering
(DIS)
in a rather simplified context,
that is we will work in $\pts$.
The scalar $\pts$ model share with QCD some important aspect. It is
an asymptotically free theory and the Feynman diagrams have the same
topology as in QCD. Moreover there are no soft but only
collinear divergences and so factorization
become simpler to deal with.
For this reasons it has been customary,
in the past, to face a QCD problem
by studying it
first
in $\pts$ \cite{taylor,kazama,kub,bfk,css}.

In $\pts$ one can define as in QCD a scalar parton
density and a fragmentation
function which obey DGLAP evolution equations \cite{kazama,css}.
It is in this framework that
we are going to calculate here the one particle deep inelastic 
cross section
in the limit $t\ll Q^2$. The aim of such a calculation is twofold.
On one hand we want to verify that the renormalized one particle
cross section gets large $\log Q^2/t$ corrections, as one
naturally expects in the two scale regime
$t\ll Q^2$. On the other hand we want to check the consistence of our
result with the factorization put forward in Ref.\cite{new}.

This paper is organized as follows: in Sect.2 we will recall the result
one gets in $\pts$ for inclusive DIS. In Sect. 3 we will perform the
calculation of the semi-inclusive cross section. In Sect. 4 we will
show how an interpretation in terms of cut vertices can be given.
In Sect. 5 we will sketch our conclusions.

\resection{DIS}

As a first step we will focus on the deep inelastic inclusive cross section.
We consider the process
$p+J(q)\to X$ where $J=\f{1}{2} \p^2$ is a scalar operator which plays the
role of the electromagnetic current.
We define as usual
\begin{equation}
Q^2=-q^2~~~~~~x=\f{Q^2}{2pq}.
\end{equation}
The structure function is
\begin{equation}
W(x,Q^2)=\f{Q^2}{2\pi} \int d^6y e^{iqy} <\!p|J(y)J(0)|p\!>.
\end{equation}
\begin{figure}[htb]
\begin{center}
\begin{tabular}{c}
\epsfxsize=5truecm
\epsffile{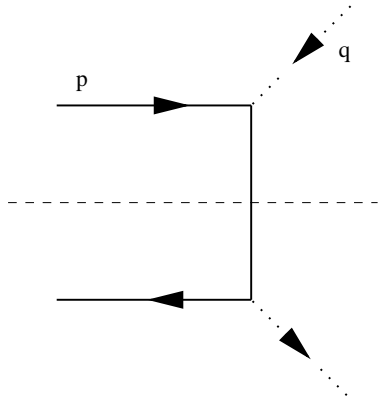}\\
\end{tabular}
\end{center}
\caption{{\em Lowest order contribution to the deep inelastic cross section}}
\end{figure}
We want to calculate the parton-current cross section $w(x,Q^2)$
at $p^2=0$ in dimensional regularization ($D=6-2\eps$), by retaining only
divergent contributions.
At Born level we have (see Fig. 1)
\begin{equation}
w_0(x,Q^2)=\f{Q^2}{2\pi} 2\pi \delta((p+q)^2)=\delta(1-x).
\end{equation}
The first order corrections are shown in Fig. 2 and give
\begin{equation}
w_{1a}(x,Q^2)=\l x(1-x)
\left( -\f{1}{\epsilon}\right)\left(\f{4\pi\mu^2}{Q^2}\right)^\epsilon
\end{equation}
\begin{equation}
\label{eq2}
w_{1b}(x,Q^2)=2\times\f{1}{2}\l \f{1}{\epsilon}\mq\delta(1-x)
\end{equation}
\begin{equation}
w_{1c}(x,Q^2)=\f{1}{12}\l \left( -\f{1}{\epsilon}\right)\mq\delta(1-x).
\end{equation}
The factor $2$ in eq. (\ref{eq2}) takes into account the
contribution from the symmetric diagram.
The diagram in Fig.2d gives only a finite correction. External self energies
have not be taken into account\footnote{
Actually one should also include self energy corrections
on the current line, which are of the same order in $\lambda$ and
cause the mixing of the operator $\phi^2$ with
the operator $\Box \phi$ \cite{collins}.
Nevertheless these diagrams are vacuum polarization of
the external probe, and in the real world are suppressed by
one power of $\alpha_{em}$ and so are neglected. We decide here
to follow the literature \cite{kub,bfk,css}
and not to include them in our
definition of the structure function.}
since we work at $p^2=0$.
\begin{figure}[htb]
\begin{center}
\begin{tabular}{c}
\epsfxsize=12truecm
\epsffile{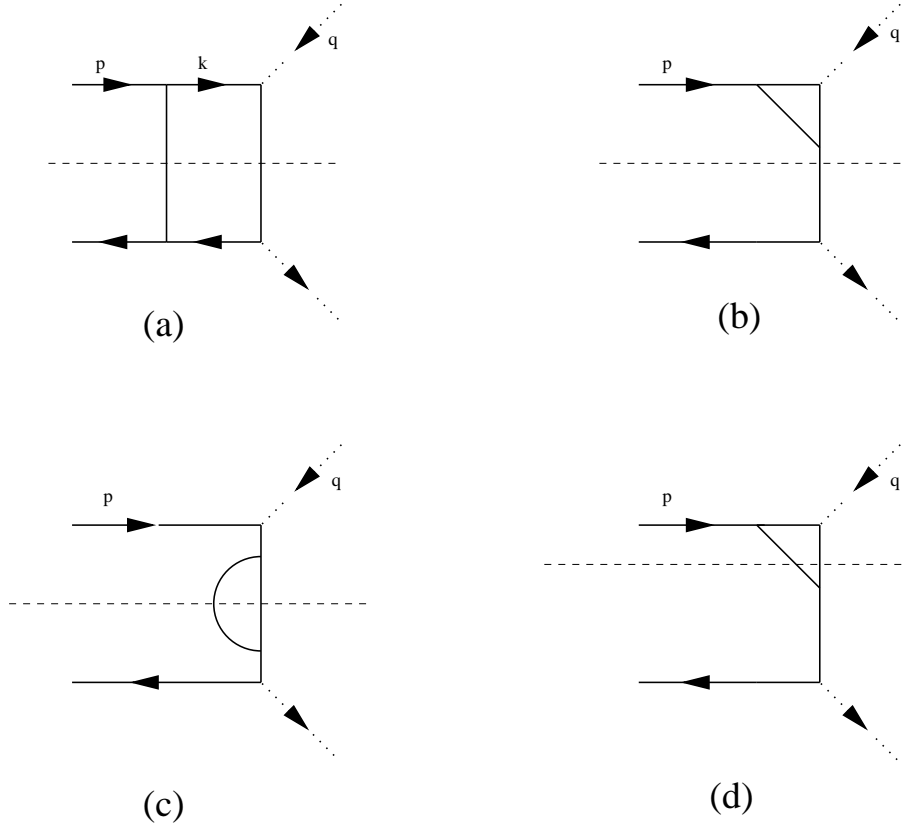}\\
\end{tabular}
\end{center}
\caption{{\em One loop corrections to the deep inelastic cross section}}
\end{figure}
Usually one has not to worry about current renormalization,
because the electromagnetic current is not renormalized by
strong interactions. However in our simplified model the operator
$J_0=\f{1}{2} \p_0^2$
is renormalized by the interaction.
The renormalization constant $Z_J$ is defined by
\begin{equation}
J=Z_J^{-1} J_0.
\end{equation}
We get, in the MS scheme
\begin{equation}
Z_J=1+\f{5}{12} \l\f{1}{\epsilon}.
\end{equation}
As a matter of fact, being the renormalization scale dependent,
the coupling to the current
becomes scale dependent, so it turns out more convenient to define a $Q^2$
dependent renormalization
\begin{equation}
Z_J(Q^2)=1+\f{5}{12} \l\f{1}{\epsilon}\mq
\end{equation}
which takes into account this effect. 
Therefore the cross section is obtained summing up all the contributions
and multiplying by $Z_J^{-2}(Q^2)$. Up to finite corrections we get
\begin{equation}
\label{ris}
w(x,Q^2)=\delta(1-x)+\l P(x)\left(-\f{1}{\eps}\right)\mq
\end{equation}
where
\begin{equation} 
P(x)=x(1-x)-\f{1}{12}\delta(1-x)
\end{equation}
is the DGLAP kernel for this model \cite{kazama,jet}.
We see that this result has the same structure one gets in QCD.
The contribution to the structure function is obtained as a convolution
with the bare parton density $f_0(x)$
\begin{equation}
W(x,Q^2)=\int_x^1 \f{du}{u}f_0(u) w(x/u,Q^2).
\end{equation}
The collinear divergence in $w(x,Q^2)$ can be absorbed as usual by
defining a $Q^2$ dependent scalar parton density $f(x,Q^2)$ by
means of the equation
\begin{equation}
\label{pd}
f_0(x)=\int_x^1 \f{du}{u} \left[\delta(1-u)+\l P(u)\f{1}{\eps}\mq\right]
f(x/u,Q^2).
\end{equation}
This renormalized scalar parton density
obeys the DGLAP evolution equation
\begin{equation}
Q^2\f{\d}{\d Q^2} f(x,Q^2)=\int_x^1 \f{du}{u} P(u) f(x/u,Q^2).
\end{equation}

On the same footing one can define for the process $J(q)\to p+X$ with
$q$ timelike a fragmentation function $d(x,Q^2)$ which obeys the same DGLAP
evolution equation. Thanks to the Gribov-Lipatov reciprocity relation \cite{gl}
at one loop level the timelike DGLAP kernel is the
same as in the spacelike case, but this relation is broken at
two loop, as explicitly verified for $\pts$ in Ref.\cite{kub}.

\resection{Semi-inclusive DIS}
We are now ready to discuss the process $p+J(q)\to p^\prime+X$.
In this case a semi-inclusive structure function can be defined as
\begin{equation}
W(p,p^\prime,q)=\f{Q^2}{2\pi} \sum_X\int d^6x e^{iqx} <\!p|J(x)|p^\prime X\!>
<\!X p^\prime|J(0)|p\!>.
\end{equation}
We define
\begin{equation}
z=\f{p^\prime q}{pq}.
\end{equation}
We will calculate the partonic cross section in the limit
\begin{equation}
\label{a} 
t\ll Q^2
\end{equation} 
at leading power, by keeping only divergent terms and possible $\log Q^2/t$
contributions. It turns out that the approximation (\ref{a}) selects a
special class of diagrams, those in which the produced particle 
is radiated by the incoming one.
As a matter of fact at the lowest order in $\lambda$ there is no contribution in the region
$t\ll Q^2$. The first diagram is the one in Fig. 3 which gives
\begin{figure}[htb]
\begin{center}
\begin{tabular}{c}
\epsfxsize=5truecm
\epsffile{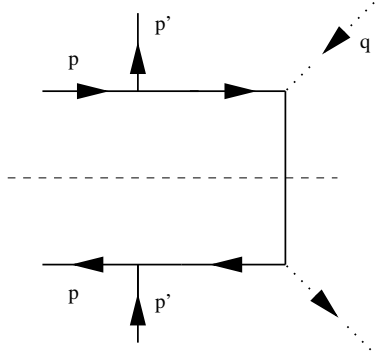}\\
\end{tabular}
\end{center}
\caption{{\em Leading order contribution to the one particle
deep inelastic cross section in the region $t\ll Q^2$}}
\end{figure}
\begin{equation}
w_1(x,z,t,Q^2)=\f{\lambda_0^2}{t^2} x \delta(1-x-z)
\end{equation}
where $\lambda_0$ is the bare coupling constant.
Notice that,
at this
order it is
necessary to distinguish between $\lambda$ and $\lambda_0$.
\begin{figure}[htb]
\begin{center}
\begin{tabular}{c}
\epsfxsize=11.7truecm
\epsffile{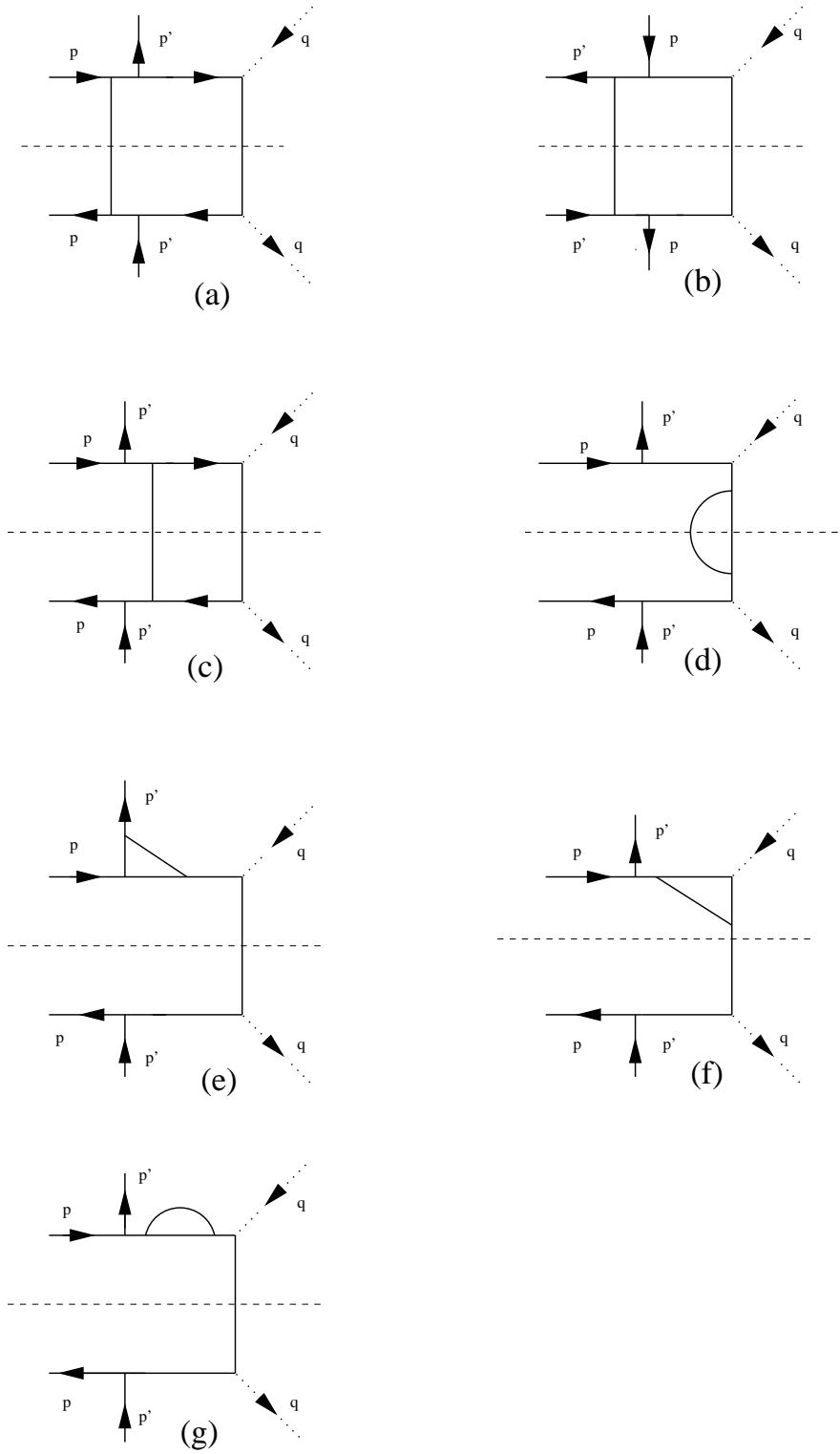}\\
\end{tabular}
\end{center}
\caption{{\em One loop leading contributions to the one particle deep inelastic cross section}}
\end{figure}

\begin{figure}[htb]
\begin{center}
\begin{tabular}{c}
\epsfxsize=12truecm
\epsffile{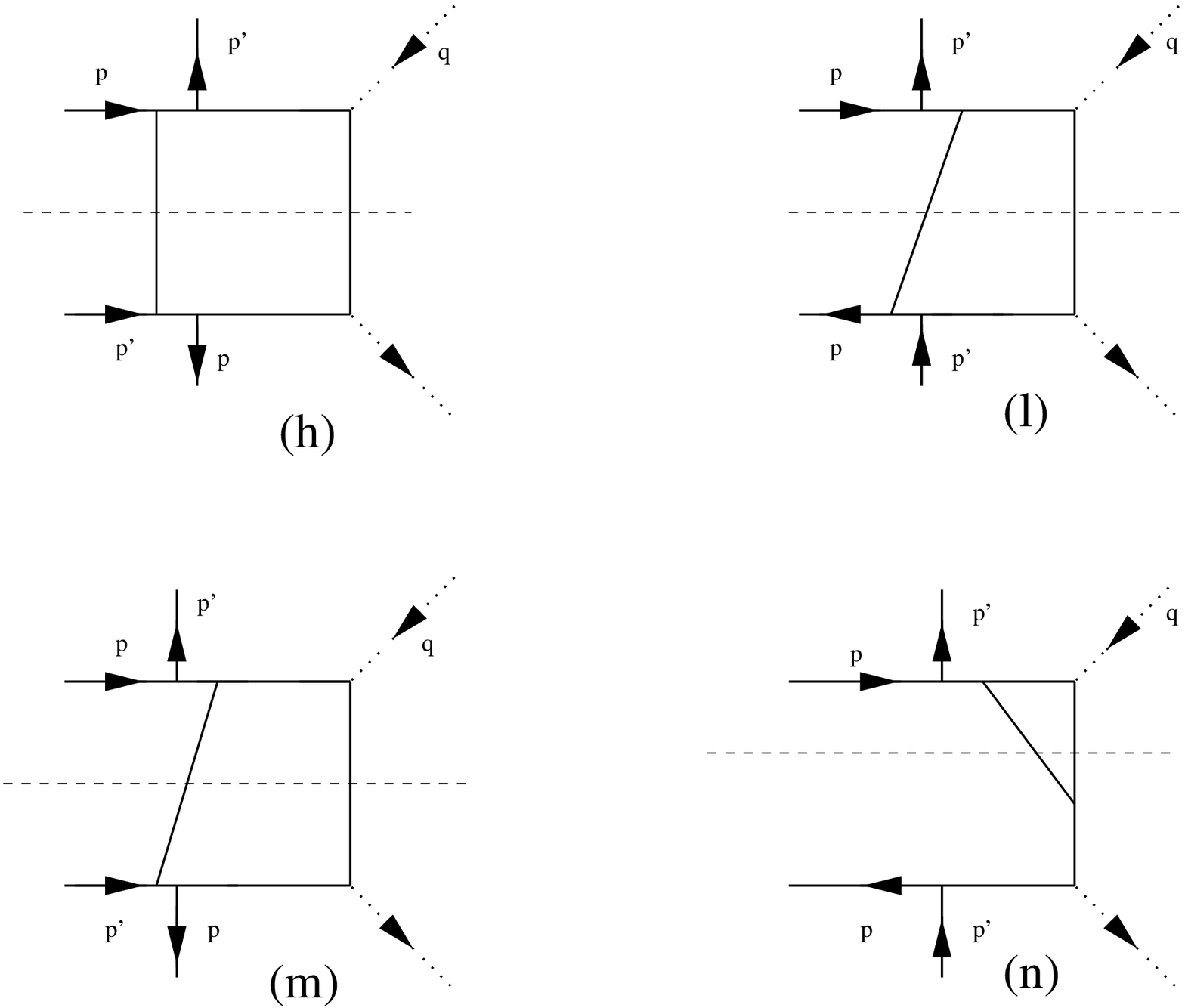}\\
\end{tabular}
\end{center}
\caption{{\em One loop finite corrections to the one particle deep inelastic cross section}}
\end{figure}

The other contributions come from the one loop corrections listed
in Figs. 4,5. It appears that the topology of the diagrams
is much richer than the one for the inclusive scattering.
The diagram in Fig. 4(a) gives
\begin{equation}
w_{2a}(x,z,t,Q^2)=\f{Q^2}{2\pi}\int \f{d^Dk}{(2\pi)^D}
\f{1}{k^4}\f{1}{(k-p^\prime)^4} \lambda^4 2\pi\delta_+\left((p-k)^2\right)
2\pi\delta_+\left((k-p^\prime+q)^2\right).
\end{equation}
We choose as a pair of lightlike vectors $p$ and $q^\prime\equiv q+xp$
and set
\begin{equation}
k=\alpha p+\beta q+\kkk
\end{equation}
\begin{equation}
p^\prime=\ap p+\beta^\prime q+\ppp
\end{equation}
\begin{equation}
s=2pq.
\end{equation}
The delta functions can be used to integrate over $\alpha$ and $\beta$.
It turns out that the integral
over $\kkk$
gets the leading contribution in the
region of small $\kkk^2$. In this limit we have
\begin{align}
&w_{2a}(x,z,t,Q^2)\simeq \f{1}{2}x(1-x-z)^3\lambda^4\nn\\
&\times\int\f{d^{D-2}\kkk}{(2\pi)^{D-1}}\f{1}{\kkk^4}
\f{1}{\left((1-z)\kkk^2-2(1-x-z)\ppp\kkk+(1-x-z)(x+z)t\right)^2}
\end{align}
so there is a collinear divergence at $\kkk=0$ which corresponds to the
configuration in which $k$ becomes parallel to $p$.
The result is
\begin{equation}
w_{2a}(x,z,t,Q^2)=\l\cc\f{1-x-z}{(x+z)^2}\left(-\f{1}{\eps}\right)\mt.
\end{equation}
In the same way the diagram in Fig. 4(b) has a collinear pole when
$k$ gets
parallel to $p^\prime$ and we get
\begin{equation}
w_{2b}(x,z,t,Q^2)=\l\cc\f{1-x-z}{(1-x)^2}\left(-\f{1}{\eps}\right)\mt.
\end{equation}
The other calculations are straightforward and give
\begin{equation}
w_{2c}(x,z,t,Q^2)=\l\f{\lambda^2}{t^2}\f{x}{1-z}(1-\f{x}{1-z})\log\f{Q^2}{t}
\end{equation}
\begin{equation}
w_{2d}(x,z,t,Q^2)=\f{1}{12}\l\cc\left(-\f{1}{\eps}\right)\mq\delta(1-x-z)
\end{equation}
\begin{equation}
\label{eqe}
w_{2e}(x,z,t,Q^2)=2\times\f{1}{2}\l\cc\left(\f{1}{\eps}\right)\mt\delta(1-x-z)
\end{equation}
\begin{equation}
w_{2f}(x,z,t,Q^2)=2\times\f{1}{2}\l\cc\left(\f{1}{\eps}\right)\mq\delta(1-x-z)
\end{equation}
\begin{equation}
\label{eqg}
w_{2g}(x,z,t,Q^2)=2\times\f{1}{12}\l\cc\left(-\f{1}{\eps}\right)\mt\delta(1-x-z).
\end{equation}

Again the factor $2$ in eqs. (\ref{eqe})-(\ref{eqg}) takes
into account the contribution of the symmetric diagrams.
The diagrams in Fig. 5 are finite and don't give
$\log Q^2/t$ contributions so they will be neglected.

There are of course many other diagrams which have not been considered here.
They are either diagrams in which the produced particle
comes from the fragmentation
of the current and diagrams which represent interference
between target and current fragmentation.
However one can verify immediately that these diagrams
are suppressed by powers of
$t/Q^2$, and so they are not relevant here.
For example, consider the diagram
depicted in Fig.6(a), in which the observed final state particle comes from
current fragmentation.
The explicit calculation shows that this diagram is suppressed by
two powers of $t/Q^2$. This fact can be better understood if we redraw
it as in Fig. 6(b): in the large $Q^2$ limit
there are actually four lines going into a hard vertex and the diagram
is suppressed by power counting \cite{new}.
\begin{figure}[htb]
\begin{center}
\begin{tabular}{c}
\epsfxsize=12truecm
\epsffile{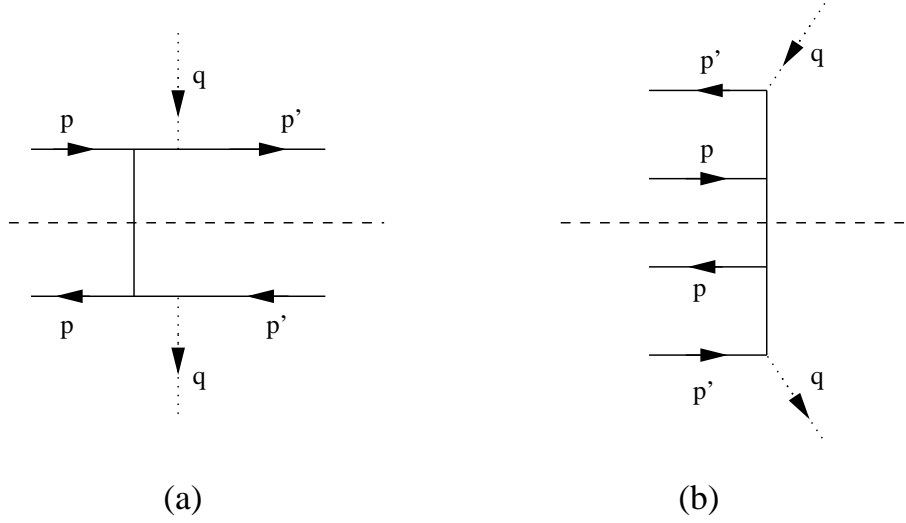}\\
\end{tabular}
\end{center}
\caption{(a) {\em A current fragmentation contribution to the 
semi-inclusive cross section}
$~~~~~~$(b) {\em The same diagram depicted in a different form}}
\end{figure}
Summing up all the contributions, multiplying by $Z_J^{-2}(Q^2)$,
introducing the running coupling constant
\begin{equation}
\lambda^2(t)=\f{\lambda^2}{1+\beta_0 \lambda^2 \log t/4\pi\mu^2}
\simeq\lambda^2\left(1-\f{3}{4}\f{1}{(4\pi)^3}\lambda^2\log\f{t}{4\pi\mu^2}\right)
\end{equation}
and using
\begin{equation}
\lambda^2_0=\lambda^2 \left(1-\f{3}{4}\l\f{1}{\eps}\right)
=\lambda^2(t) \left(1-\f{3}{4}\l\f{1}{\eps}\mt\right) 
\end{equation}
we finally get
\begin{align}
\label{sris}
w(x,z,t,Q^2)&=\f{\lambda^2(t)}{t^2}x
\Big(\delta(1-x-z)+\f{1}{6}\delta(1-x-z)\l\f{1}{\eps}\mt\nn\\
&+\l\left(\f{1-x-z}{(x+z)^2}+\f{1-x-z}{(1-x)^2}\right)
\left(-\f{1}{\eps}\right)\mt\nn\\
&+\f{1}{x} P\left(\f{x}{1-z}\right)\l\log\f{Q^2}{t}\Big).
\end{align}
We have now to verify that the collinear divergences in eq. (\ref{sris})
can be correctly absorbed in the redefinition of parton density
and fragmentation function.
We have
\begin{align}
W(x,z,t,Q^2)&=\int_0^1 \f{du}{u}\int_0^1\f{dv}{v^4}
f_0(u)w(up,p^\prime/v,q)d_0(v)\nn\\
&=\int_0^1 \f{du}{u}\int_0^1\f{dv}{v^4}
f_0(u)w(x/u,z/uv,tu/v,Q^2)d_0(v).
\end{align}
Here the integration measure $dv/v^4$ takes into account the scaling of
the phase space element of the detected particle.

By using the definition of scale dependent parton density (\ref{pd})
and the corresponding definition for fragmentation function,
eventually we find
\begin{align}
\label{final}
W(x,z,t,Q^2)&=\int_{x+z}^1 \f{du}{u}\int_{\f{z}{u-x}}^1\f{dv}{v^4}
f(u,t)\f{\lambda^2(t)}{t^2}\f{v^2}{u^2}
\Big[\delta\left(1-\f{x/u}{1-z/uv}\right)\nn\\
&+\l P\left(\f{x/u}{1-z/uv}\right)\log\f{Q^2}{t}\Big] d(v,t)
\end{align}
where again only leading $\log Q^2/t$  terms have been considered
and the integration limits are derived using
momentum conservation.

Eq. (\ref{final}) is our final result.
We see that the renormalized hard cross section gets, as expected, a
large $\log Q^2/t $ corrections whose coefficient is precisely
the DGLAP kernel.
Such correction, if not properly resummed,
can spoil perturbative calculations in the region $t\ll Q^2$.

Moreover we find that in the limit $t\to 0$ a new singularity
appears in the cross section, which corresponds to the configuration
in which $p$ becomes parallel to $p^\prime$. As pointed out
in Ref.\cite{grau}, when integrating over $t$,
such singularity can't be absorbed in ordinary parton densities
and fragmentation functions and  the introduction
of a new phenomenological distribution, the fracture function \cite{tv}
becomes necessary.

Eq. (\ref{final}) can also be rewritten in the following form
(see Fig. 7):
\begin{figure}[htb]
\begin{center}
\begin{tabular}{c}
\epsfxsize=7truecm
\epsffile{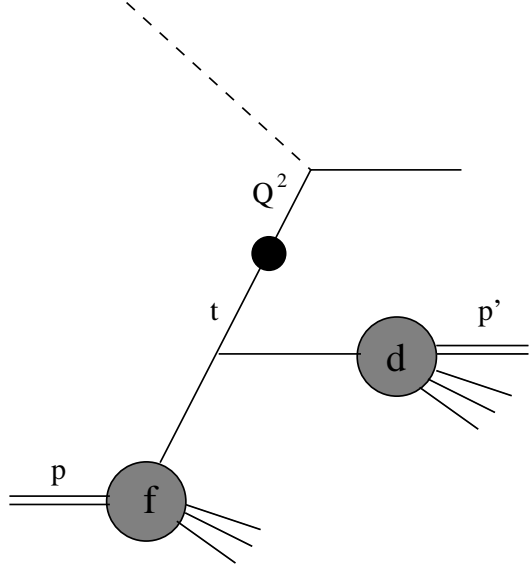}\\
\end{tabular}
\end{center}
\caption{{\em Graphical interpretation of eq. (\ref{eqjet})}}
\vspace*{1cm}
\end{figure}
\begin{align}
\label{eqjet}
W(x,z,t,Q^2)&=\f{\lambda^2(t)}{zt^2}
\int_x^{1-z}\f{dr}{r}\int_{z+r}^1 \f{du}{u(u-r)} {\hat P}
\left(\f{r}{u}\right) f(u,t)\Big[\delta\left(1-\f{x}{r}\right)\nn\\
&+\l P\left(\f{x}{r}\right)\log\f{Q^2}{t}\Big] d\left(\f{z}{u-r},t\right)
\end{align}
where we have defined the A-P real scalar vertex ${\hat P}(x)=x(1-x)$.
The function
\begin{equation}
E^{(1)}(x,Q^2,Q^2_0)=\delta(1-x)+\l P(x) \log \f{Q^2}{Q^2_0}
\end{equation}
is the first order approximation of the evolution kernel $E(x,Q^2,Q^2_0)$
which resums the leading logarithmic series \cite{jet}.
This fact suggests that
an interpretation of eq. (\ref{eqjet}) can be given in terms of
Jet Calculus \cite{jet}.

\resection{Interpretation in terms of cut vertices}

The cut vertex expansion is a generalization
of the Wilson expansion originally
proposed by Mueller in Ref.\cite{mueller} and applied
to a variety of hard processes in Ref.\cite{gupta}.
We will briefly recall it for DIS in $\pts$. 

Let us go back to Sect.2 and
set $p^2<0$ with $p=(p_+,{\bf 0},p_-)$.
If we choose a frame in which $p_+\gg p_-$ we can write 
for the parton-current cross section \cite{mueller}
\begin{equation}
w(p,q)=\int \f{du}{u} v(p^2,u)C(x/u,Q^2)
\end{equation} 
where
\begin{equation}
\label{cut}
v(p^2,x)=\int V(p,k)x\delta\left(x-\f{k_+}{p_+}\right)\f{d^6k}{(2\pi)^6}
\end{equation}
is a spacelike cut vertex,
which fully contains the mass singularities of the cross section,
and
$C(x,Q^2)$ is the corresponding coefficient function. In eq. (\ref{cut})
$V(p,k)$ is
defined as
the discontinuity of the four point amplitude in the channel
$(p-k)^2$ and the integration over $k$ is properly renormalized.

The results of the previous sections can be easily recast in the
form of cut vertex expansion. If we define
\begin{equation}
v(x,\eps)=\delta(1-x)+\l P(x)\left(-\f{1}{\eps}\right)
\end{equation}
\begin{equation}
C(x,Q^2)=\delta(1-x)+\l P(x) \log\f{Q^2}{4\pi\mu^2}
\end{equation}
we can write eq. (\ref{ris}) in the form
\begin{equation}
w(x,Q^2)=\int_x^1 \f{du}{u} v(u,\eps)C(x/u,Q^2).
\end{equation}
Here $v(x,\eps)$ is a spacelike cut vertex defined at $p^2=0$ whose mass divergence is regularized dimensionally.

\begin{figure}[htb]
\begin{center}
\begin{tabular}{c}
\epsfxsize=14truecm
\epsffile{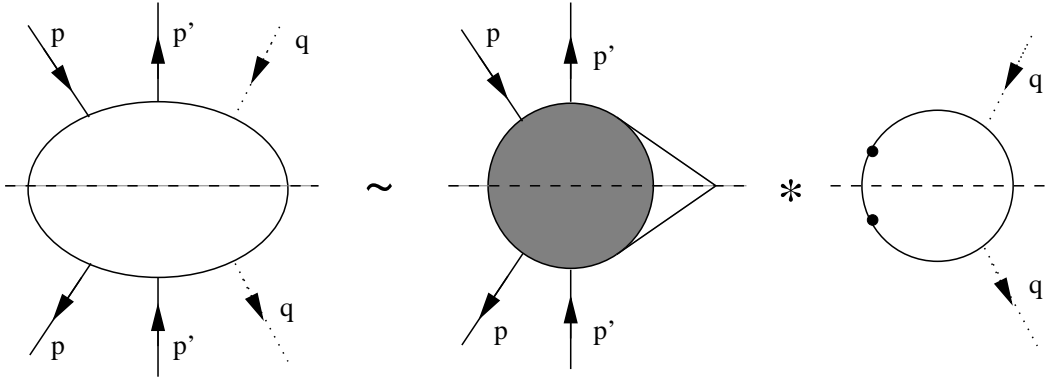}\\
\end{tabular}
\end{center}
\caption{{\em Factorization of the semi-inclusive structure function}}
\end{figure}

It turns out that
a similar interpretation can be given of eq. (\ref{sris}).
We define
\begin{equation}
\xb=\f{x}{1-z}
\end{equation}
and
\begin{align}
v(\xb,z,t,\eps)&=\f{\lambda^2(t)}{t^2}\Big[\delta(1-\xb)
+P(\xb)\l\log\f{4\pi\mu^2}{t}+\l\f{1}{\eps}\mt\nn\\
&\times\left(\f{1}{6}\delta(1-\xb)
-\f{(1-z)^2\xb(1-\xb)}{(\xb(1-z)+z)^2}
-\f{(1-z)^2\xb(1-\xb)}{(1-\xb(1-z))^2}\right)\Big]
\end{align}

as a {\em generalized cut vertex} \cite{new} which contains all the leading
mass singularities of the cross section.
We can write up to ${\cal O}(t/Q^2)$ corrections (see Fig. 8)
\begin{equation}
\label{fact}
w(\xb,z,t,Q^2,\eps)=\int_\xb^1 \f{du}{u}
v(u,z,t,\eps)C(\xb/u,Q^2)
\end{equation}
where the coefficient function is the same which occurs in
inclusive DIS.

It is well known that in dimensional regularization there is a mixing between
collinear and ultraviolet divergences. In order to avoid it one should
distinguish between
$\eps$ and $\eps^\prime$ to regulate UV and 
collinear divergences, respectively.
Moreover external self energies should be taken into account
since they are zero on shell for a cancelation of the two kind of divergences.
In this framework one can show that factorization (\ref{fact}) actually
holds graph by graph \cite{graz}.

\resection{Conclusions}

In this paper we have studied the deep inelastic semi-inclusive cross section
in the target fragmentation region and we have performed an explicit calculation in $\pts$ model field theory. We have shown that the renormalized hard cross section gets a large $\log Q^2/t$ correction as expected in a two scale regime.
Furthermore we have found that the coefficient driving this logarithmic correction is precisely the scalar DGLAP kernel. This result suggests that the $Q^2$ dependence of the cross section
in such processes at fixed $z$ and $t$
is driven by the same anomalous dimension which controls the inclusive DIS,
as proposed in Ref. \cite{new}, and in
in the context of diffraction in Ref.\cite{diff}.

We have then examined our result from the point of view of extended factorization and we have found that it is
consistent with such an hypothesis.
In this framework
the partonic semi-inclusive cross section factorizes into a convolution of a
new object, a generalized cut vertex $v(p,p^\prime,\xb)$ \cite{new},
with four rather than two external legs, and a coefficient function $C(\xb,Q^2)$. The former is of long distance nature
and embodies the leading mass singularities of the cross section, while the latter is of perturbative nature and, what is important,
it is the same as in inclusive DIS.

Therefore these results verify the validity of the approach proposed in
Ref. \cite{new}. Of course the calculation performed here is only a one
loop calculation in a scalar model. Nevertheless we believe
that the
results
obtained
maintain
the same structure
in QCD, by assuming
in $\pts$ a particularly simple and appealing form.
\vspace{3mm}
\begin{center}
\begin{large}
{\bf Acknowledgments}
\end{large} 
\end{center}
\noindent The author would like to thank
G. Camici, S. Catani, D.E. Soper and
G. Veneziano for useful discussions, and particularly L. Trentadue for
his advice and encouragement during the course of this work.

\end{document}